----------
X-Sun-Data-Type: default
X-Sun-Data-Description: default
X-Sun-Data-Name: prlpol.tex
X-Sun-Charset: us-ascii
X-Sun-Content-Lines: 686

\documentstyle[aps,multicol] {revtex}
\renewcommand{\narrowtext}{\begin{multicols}{2} \global\columnwidth20.5pc}
\renewcommand{\widetext}{\end{multicols} \global\columnwidth42.5pc}
\newcommand{\q}{{\bf q}}

\begin{document}
\draft

\title{
Splitting of  Landau levels of a 2D electron due to electron-phonon interactions
}

\author{V. Cheianov}

\address{Uppsala University, S-751 08, Uppsala, Sweden}
\author{A.P. Dmitriev  and
V.Yu.Kachorovskii }
\address{
A.F.Ioffe Physical-Technical Institute, St.Petersburg, 194021,
Russia}

\maketitle

\begin{abstract}

We show that in a very strong magnetic field $B$ electron-phonon
interaction gives rise to a splitting of Landau
levels
of a 2D electron into a series of infinitely
degenerate sublevels. We provide both qualitative and quantitative
description of this phenomenon. The cases of interaction
with acoustic and polar optical phonons are considered. The
energy distance between nearest sublevels  in
both cases tends to  zero as $B^{-1/2}$ at large $B$.

\end{abstract}

\narrowtext

The dynamics of a 2D lowest Landau level (LL) electron
in high mobility samples where only the electron-phonon
interaction is relevant was disscussed in Refs.\onlinecite{ior,mi}.
In the limit of weak electron-phonon coupling
the temperature dependence of the longitudinal electron
mobility was calculated in these works both at low \cite{ior}
and relatively high \cite{mi} temperatures.

 The limiting case of
weak electron-phonon interaction is not, however, realized
at very high magnetic fields.
The effective
coupling constant $\alpha$  characterizing the electron-phonon
interaction in a strong magnetic field increases with $B$ (see below) and
may reach large values at which a
new physics comes into play due to the essential role of
polaronic effects. In this limit the
conclusions of Refs.\onlinecite{{ior},{mi}} are no longer
valid.
Some aspects of the problem of calculating the
electron mobility in this case were discussed in
Ref. \onlinecite{kuk}.

In this letter we would like to concentrate on a
simpler problem of finding the
energy spectrum of the
electron-phonon system when
$ B \rightarrow \infty$. Knowledge
of this spectrum  is nesesary for the
correct calculation of the electron mobility at
very low temperatures.

For $\alpha \gg 1$ the lattice in the vicinity of an electron
is strongly
deformed and this deformation creates a polaronic well
(PW) around the electron.
The characteristic frequency of the electron
motion in the PW is very
high compared to the phonon frequencies and the lattice deformation
forming the PW can be
considered quasistatic. The
energy and the wave function of the ground state (GS) of the electron
in the PW in the limit $ B \rightarrow \infty$  were
calculated in Ref.\onlinecite{kuk1} in the framework of
Pekar adiabatic method \cite{pek}.
At first sight it may seem that the results of Ref. \onlinecite{kuk1}
give the complete description of the physics of the problem
and the polaronic effects simply reduce
to an energy shift of the LL. This, in fact, is not true.
As we will show, in strong coupling limit the
LL splits into a series of sublevels, these
sublevels being infinitely degenerate like the original LL.

This splitting can be understood qualitatively on an example
of a simple model, which can be solved exactly.
Let us consider a 2D electron with the effective mass
$m^*$ placed in a strong magnetic field and coupled by a
harmonic potential to a heavy neutral particle with the
mass $M \gg m^*$. This particle models the phonon cloud
surrounding an electron in the crystall.  There are three different types of motion in the
system.
The fastest one is the
motion of the electron with the cyclotronic frequency
$\omega_c = eB/m^* c $. The second one is a
slower drift of the magnetic circle center of the electron
around the heavy particle with the frequency
 $ \Omega= \omega_0^2 /\omega_c$, where
$\omega_0 \ll \omega_c$ is the frequency of the
electron oscillations in the harmonic potential in the
absence of the magnetic field. The third,
slowest type of motion is the cyclotronic
rotation of the system as a whole with the frequency
$\Omega_c = eB/Mc $. The energy spectrum
of the model can be classified by three quantum numbers
$n,m, N$ corresponding to these types of motion $E_{n,m,N } =
\hbar\omega_c n + \hbar \Omega m + \hbar\Omega_c N $.

In case of an electron interacting with
phonon field there also exist three types of motion. In
analogy with the model, the Landau levels
corresponding to the fast cyclotronic rotation of the electron
are split due to the slower drift of the magnetic circle
center in the PW and to yet slower rotation of the polaron
as a whole. In this letter we restrict ourselves to the calculation of
the splitting of the lowest LL. As we  will  show the drift frequency
$\Omega$ depends on the radius  of the drift orbit and
the series associated with the quantum number $m$ is not
equidistant. The frequency of the slowest cyclotronic rotation
of the polaron as a whole $\Omega_c$ is inversely
proportional to the polaronic mass $M$, which depends on
the magnetic field and the quantum number $m$. The
appearance of a finite mass of the polaron (an electron at
the LL has effectively no mass), is a result of dressing of the electron
by phonons. We want to emphasize that in our considerations
the transitions between different LLs of
the electron are neglected and the finite polaronic
mass arises when the states of one LL only are taken into account.
In strong coupling limit the mass $M$ is proportional
to $\alpha$ and grows with magnetic field as $ B^{3/2}$, whereas
the distance between nearast sublevels $\hbar \Omega_c $
decreases as $ B^{-1/2}$.
Note also, that all the sublevels of the spectrum obtained in our letter
are infinitely degenerate over the position of the
magnetic circle center of the polaron as a whole.

We consider an electron at zero temperature in a quantum
well (QW) of the width $a$
located in the
plane $(x,y)$ to which a strong magnetic field $\bf B$
parallel to the axis $z$ is applied.  We assume that
the electron is at
the first level  in QW
and at the lowest LL.
The sample is supposed to be fairly pure and the filling
factor of the LL is supposed to be small so that the impurity
scattering as well as the electron-electron scattering can be neglected.
The Hamiltonian of the system projected on the lowest
LL can be written \cite{{mi},{stat.wave}} in terms of the
operator of magnetic circle center $ {\bf R}=(\hat X, \hat Y) $ with
non-commuting components $[\hat X, \hat Y]=i l^2$:
\begin{equation}
H=\sum_{\bf q} \frac{ P_{\bf q}^2+\omega_{\bf q}^2 Q_{\bf q
}^2}{2}+
\sum_{\bf q}  f_{\bf q}
Q_{\bf q} \sin({\bf q}_{||}{\bf R}-{\pi}/{4}),
\label{Ham}
\end{equation}
where $l$ is the magnetic length,  $ Q_{\bf q}$
are the normal coordinates of the oscillations of the
lattice and $P_{\bf q }$ are the momenta conjugated to
them, $\omega_{\bf q} $ is the phonon frequency, ${\bf q}$
is the phonon wave vector with the components
$ {\bf q}_{||}$ in the QW's plane and $q_z$ in $z$
direction.
The function $f_\q$ can be written as
\begin{equation}
 f_\q= F_\q  \eta(q_z) e^{-{q_{||}^2l^2}/{4}}, \:\:
 \eta(q_z)=\int d z e^{i q_z z} |\psi_0(z)|^2.
\label{fq}
\end{equation}
The factors $ \exp(-{q_{||}^2l^2}/{4})$ and $\eta(q_z) $
%
%
appear as a result of the projecting of the
original Hamiltonian of the system on the zero LL and on
the lowest level in the QW with the
wave function of transversal quantization $\psi_0(z)$.
Due to these factors the electron effectively interacts
with the phonon modes with momenta lying in the shell
$ q_{||}<1/l,\ q_z<1/a$.
The function $ F_\q$ is proportional to the square root
of coupling constant $\alpha$.
Its concrete form
depends on the type of phonons. We will do all the
calculations for arbitrary $F_\q$ and afterwards consider
two examples of polar optical phonons and acoustic
phonons interacting with the electron via deformation potential.

 We
assume that at $\alpha \gg 1 $ characterisic frequency of electron motion
in the PW is much larger than phonon
frequencies. Then the PW is quasistatic and
one can introduce the wave function $\Psi$ of the GS of the
electron in this well.  Averaging the
Hamiltonian (\ref{Ham}) with $\Psi$ we get the
Hamiltonian which only depends on the phonon coordinates and
momenta. The ground state  energy of this
Hamiltonian  is given by

\begin{equation}
\label{Enerfunct}
E\{ \Psi \} = - \sum_{\bf q}
\frac{ f_{\bf q}^2}{2\omega_\q^2}  \langle \Psi|
\sin({\bf q}_{||} {\bf R}-{\pi}/{4})
|\Psi\rangle^2.
\end{equation}
In Pekar method \cite{pek} $\Psi$  should be obtained
by minimizing the energy functional (\ref{Enerfunct}) under
the constraint $ \langle \Psi|\Psi\rangle = 1 $.
One can see that the functional (\ref{Enerfunct})
is stationary on the eigenfunctions of the angular momentum
operator of the electron at the lowest LL
$\:\: \hat M =\hbar
[l^2-\hat X^2-\hat Y^2]/2l^2.$
This implies that the self consistent potential of the
PW is axially symmetric.
The absolute minimum of
the energy functional (\ref{Enerfunct}) is reached on the zero angular
momentum wave function \cite{kuk1} which  in $X$ representation
( $\hat X = X $ and $ \hat Y= -il^2 d/dX $)
has the form
$
\Psi_0(X)=(\pi l^2)^{-1/4} \exp(-X^2/2
l^2) .
$
The GS energy   is given by
\begin{equation}
E_0=-\sum_{\bf q} \frac{f_{\bf q}^2}{4 \omega_{\bf q}^2}
\exp\left( -{q_{||}^2 l^2}/{2} \right).
\label{Eg}
\end{equation}

Now let us consider the excited polaronic states in the
quasiclassical approximation.
The Lagrange function corresponding to the
Hamiltonian (\ref{Ham}) is written as follows
\begin{equation}
{\cal L}= \frac{\hbar}{\l^2} \dot X Y+
\sum_\q \frac{\dot Q_\q^2 - \omega_\q^2 Q_\q^2}{2}
- \sum_\q f_\q Q_\q \sin(\q_{||} {\bf R} -{\pi}/{4}).
\label{Lagr}
\end{equation}
The  equations of motion given by this Lagrange
function read
\begin{eqnarray}
\label{equaR}
\dot {\bf R} = -\frac{l^2}{\hbar}
\sum_\q & &
f_\q
Q_\q [{\bf n}\times {\bf q}_{||}]
\cos({\bf q}_{||} {\bf R}-{\pi}/{4}),
\\
\ddot Q_\q+\omega_\q^2 Q_\q & &  = - f_\q
\sin({\bf
q}_{||} {\bf R}-{\pi}/{4}),
\label{equaQ}
\end{eqnarray}
where $ \bf n$
is the unit vector in $z$ direction.
Classical motion
of the electron in the axially symmetric PW  is simply rotation
\begin{equation}
{\bf R}= {\bf R_0}+{\bf r}= (R_0\cos(\Omega t)+x,
\ R_0 \sin(\Omega t)+y),
\label{rotation}
\end{equation}
where $\bf  r$ is the position of the PW.
After substituting (\ref{rotation}) into (\ref{equaQ}) we see
that the phonon coordinates $ Q_\q $ can be represented as a sum
of harmonics with the frequencies which are integer multiples of
$ \Omega $.  Assuming that  $\Omega \gg
 \omega_\q $ we can keep in the sum only the harmonic with zero frequency and
obtain the following expression for $Q_ \q$
\begin{equation}
Q_\q \simeq Q_\q^0({\bf r}) = - \frac {J_0( q_{||} R_0) f_{\q} }{
\omega_\q^2}\sin(\q_{||} {\bf r}-{\pi}/{4}).
 \label{equqq}
\end{equation}
Substituting Eqs. (\ref{rotation}) and (\ref{equqq})
into (\ref{equaR}) we get the
frequency $\Omega$ of the classical drift of the electron in the
effective phonon potential $U(R_0)$
\begin{equation}
\Omega=\frac{l^2}{\hbar R_0} \frac{d U(R_0)}{d R_0},\:\:\:\:
 U(R_0) =-\sum_{\bf q} \frac{f_\q^2 J_0^2(q_{||} R_0)}
{4\omega_\q^2}.
\label{Dispereq}
\end{equation}
Quasiclassically the values of the rotation radius $ R_0$ are
quantized as
\begin{equation}
R_{0m}=l\sqrt{2m},\  m \gg 1
\label{Rm},
\end{equation}
where $ m$ is an absolute value of the quantum of the
angular momentum. In our
case $\:m\:$ coinsides with the number of the energy level in the  PW.
The energy of the $m$-th  high excited state is given by
\begin{equation}
E_{m}= U(R_{0m})
\label{Em}
\end{equation}
These levels are separated by large energy $ \sim \hbar \Omega $
proportional to the coupling constant.

The  high frequency motion
described above is connected with the
internal degrees of freedom of the polaron.  Besides these internal
degrees of freedom there exist two more degrees of
freedom $ x,y $ (see Eq.(\ref{rotation})) corresponding to
the motion of the polaron as a whole.
Now we will show that polaron
aquires a finite mass $ M \sim {\alpha}$ and
moves as a heavy particle with the electric charge $e$ in
magnetic field $B$. Quantization of this motion leads to
additional equidistant splitting of  energy
levels. The energy of splitting is equal to the cyclotronic
energy $ \hbar \Omega_c = \hbar eB/Mc$.

First we consider the problem quasiclassically,
assuming that the electron in the PW is in one of
the highly
exited states ( $m \gg 1$) .
Let us treat vector $\bf r $ in (\ref{rotation}) as a slowly
changing function of time. Then the  deformation field
$ Q_q^0$
will adiabatically follow the motion of the polaron according
to Eq. (\ref{equqq}). Let us do the following change of
variables
$Q_q=Q_q^0(\bf r)+\theta_q$,
where $\theta_q$ stands for the oscillations of the phonon
coordinates around  $Q_q^0$.
Substituting Eqs.(\ref{rotation}) into
(\ref{Lagr}) and averaging
over the fast rotation with the ferquency $\Omega$ we get the
effective Lagrange function corresponding to the slow motion of
the polaron

\begin{equation}
{ \cal L_{\rm ef} }
=
\frac{\hbar}{l^2}y\dot x +
\frac{M_m \dot {\bf r}^2}{2} +
\sum_{\q} \frac{\dot \theta_\q^2-\omega_\q^2
\theta_\q^2}{2} + {\cal L}_{\rm int},
\label{Efflagr}
\end{equation}
The first two terms in Lagrange function (\ref{Efflagr})
describe the motion of a massive particle  with the electric
charge $e$ in the magnetic field $B$. Classically this motion
is cyclotronic rotation with the frequency $\Omega_c $.
 The
mass $M_m$ of the polaron in $m$-th exited state is given by
\begin{equation}
M_m =\sum_\q
\frac {f_{\bf q}^2 J_0^2(q_{||} R_{0m})} {4 \omega_{q}^4} q_{||}^2.
\label{Mass}
\end{equation}
The term ${\cal L}_{\rm int} $ in the
Lagrange function describes
the
 interaction  of polaron with the phonons $\theta_\q $.
This interaction leads to decay of the cyclotronic rotation of
the polaron and $\Omega_c$ acquires  imaginary part $\Gamma$.
For acoustic phonons $\Gamma$ is small as
compared to $\Omega_c$ if the velocity of the cyclotronic motion
is less than the sound velocity.
For optical phonons the decay appears as long as we take into
account the dispersion of phonon frequency. For weak dispersion
$\Gamma$ is also small.
In both cases the quantization of cyclotronic rotation of the
polaron leads to well defined Landau levels of the polaron as
a whole.

Quasiclassical approximation is not valid if we consider
the cyclotronic splitting of the GS.
Still in this case it is also possible to derive the effective
Lagrange function for the slow motion of the polaron.
 This can be
done using the standard procedure of integrating out fast
components $ {\bf R_0}(t) $ of the vector $ {\bf R}(t) $ in
functional integral
describing transition amplitudes of the electron interacting
with the phonon field. The resulting effective Lagrange function
will only differ from (\ref{Efflagr}) by concrete form of the
interaction term and by the expression for the mass $M$. Namely,
in Eq. (\ref{Mass}) the Bessel function
$ {\rm J}_0(q_{||} R_{0m})$, which had appeared as a result of
classical averaging over fast rotation should be replaced by the
quantum average of $ \exp(i \q_{||} {\bf R})$ over the GS
of the electron $\:\exp(iyX/l^2) \Psi_0(X-x)$ in the PW
placed at the point ($x,y$). For the mass of the polaron
in the GS we have
\begin{equation}
M_0 =\sum_\q
\frac
{f_{\bf q}^2  q_{||}^2}
{4 \omega_{q}^4}
\exp\left({-\frac{q_{||}^2 l^2}{2}}\right).
\label{Mgr}
\end{equation}

The highly exited
states of an electron in the PW can decay
into lower exited states, the energy difference being transfered to
the polaron as a whole. The rigorous calculation of
corresponding transition probability is cumbersome.
One can argue, however, that this probability is negligibly
small. Indeed, the momentum $P$ transfered to polaron in this
process is $ P= \sqrt{2M(E_m -E_{m'})} \sim{ \sqrt{M \hbar \Omega}}
\sim {\alpha} $. In strong coupling limit $P$ is  large
compared to $\hbar/l$ and cannot be compensated by phonon
emission due to the factor $ \exp(-{q_{||}^2l^2}/{4})$ in Eq.(\ref{fq}).

The equations obtained above are general and can
be applied to any type of electron-phonon interaction.
Next we
consider two concrete examples of an electron
interacting with acoustic and polar optical phonons.
For certainity we will assume that
$ a \ll  l $.

1. $ \bf{ Acoustic \:\:  phonons}.$
For acoustic phonons
\begin{equation}
F_q = \sqrt{\frac{2 \pi \alpha \hbar l^2 s \omega_{q}^2}{ V}},
\:\:\: \alpha = \frac{C_0^2}{\pi \rho \hbar l^2 s^3}.
\label{Facust}
\end{equation}
Here $\alpha \sim{B}$ is the dimensionless coupling constant
characterizing the electron-phonon interaction in a strong
magnetic field ,
 $C_0$ is the deformation potential constant, $\rho$
is the density of the crystal, $s$ is the sound velosity,
and $V$ is the volume of the crystall.

Performing the summation in
Eqs.(\ref{Eg})
we obtain  energy of the GS of the polaron

\begin{equation}
E_0 = -\frac{\alpha \hbar \omega_a}{8 \pi}, \:\:
\omega_a= \frac{s}{2} \int_{-\infty}^{\infty} d q_z
\eta^2(q_z) \simeq \frac{s}{a}
\label{ac0}
\end{equation}

Classical frequency of rotation of the electron in the PW
is determined by Eq.(\ref{Dispereq} ). Taking into account
that in Eq.(\ref{Dispereq} )
$ R_0 \gg l$ we obtain the following expression for the
drift frequency

\begin{equation}
\Omega =\frac{\alpha \omega_a}{\pi
\sqrt{32\pi}}
\left(\frac{l}{R_0}\right)^3.
\label{Omegac}
\end{equation}
Eq. (\ref{Dispereq}) was derived under the assumption that
$ \Omega \gg \omega_q $. The phonon frequencies are restricted
by $ \omega_a$. Taking into account Eq. (\ref{Rm}) we have
the following condition of the
validity of our calculations
\begin{equation}
\alpha  \gg1 ,\:\:\:\:  m \ll  \alpha^{2/3}.
\label{acon}
\end{equation}

By the virtue of (\ref{Dispereq}) and (\ref{Em})
for the energy  levels in quasiclassical
approximation we have

\begin{equation}
E_{m}= - \frac{\alpha \hbar \omega_a}
{8 \pi^{3/2}\sqrt{m}}.
\label{acen}
\end{equation}

The mass of the polaron in ground and highly
excited states are obtained from Eqs.(\ref{Mass},\ref{Mgr})

\begin{equation}
M_0 = \alpha \frac{ \sqrt{\pi}\hbar}{32 l s}, \
M_{m} = \alpha \frac{\hbar}{8\pi l s\sqrt{2m} }.
\label{mac}
\end{equation}
 The spacing between Landau levels of the polaron depends on
 $m$   and is  equal to $ \hbar \Omega^0_c
=\hbar e B/(M_0 c)\ $   and   $ \hbar \Omega^{m}_c
=\hbar e B/(M_m c)\ $ for ground and highly exited states
correspondingly.   Viewing the term $ {\cal L}_{\rm int}$ in
Eq.(\ref{Efflagr}) as a perturbation one can  get the following result
for the imaginary parts $\Gamma_0 $ and $\Gamma_m$ of the
cyclotronic frequencies $\Omega^0_c$ and $\Omega^{m}_c$
\begin{equation}
\frac{\Gamma_0}{\Omega^0_c } =
\frac{8\pi}{3}\left(\frac{32}{\pi \alpha}\right)^3 , \
\frac{\Gamma_m}{ \Omega^{m}_c } =
\frac{8 \pi m^2}{3} \left(\frac{8\pi}{\alpha}\right)^3
\label{gamac}
\end{equation}
to the first
order in small parameter $ 1/ \alpha^3 $.
We see that in a strong
magnetic field when $\alpha$ is large $\Gamma$ is small
compared to the spacing $ \Omega_c$ and Landau levels of
the polaron are well defined.
Eqs.(\ref{gamac}) are valid if
the polaron velosity is smaller than the sound
velosity . In the opposite case $\Gamma$
becomes very large and cyclotronic quantization of polaron
motion is absent.

2. $\bf {Optical\:\: phonons}.$
For optical phonons
\begin{equation}
F_q = \sqrt{ \frac
{8 \alpha l \hbar\omega_0^3}
{q^2 V}},\:\:\: \alpha=\frac{\pi e^2}{\hbar \omega_0 l}
\left(\frac{1}{\epsilon_\infty}-\frac{1}{\epsilon_0}\right).
\label{Fopt}
\end{equation}
Here $\alpha \sim{\sqrt{B}}$
is the dimensionless coupling constant for optical phonons in a strong
magnetic field and
$\omega_0$ is the optical phonon frequency .
The expressions for the drift frequency and the energy of
an electron and
the  mass of the polaron in ground and highly
exited  states read
\begin{equation}
 \Omega =\frac{\alpha \omega_0}{2\pi^2}
\frac{l^3\ln(R_0/l)}{R_0^3},
\end{equation}
\begin{equation}
E_0 = -\frac{\alpha\hbar \omega_0}{ \sqrt{8\pi}}
, \:\:\:
E_m= - \frac{\alpha \hbar \omega_0}{4\pi^2}
\frac{\ln m}{\sqrt{2m}}.
\end{equation}
\begin{equation}
M_0 = \frac{\alpha \hbar}{8\sqrt{\pi} \omega_0 l^2}  , \:\:\:
M_m =\frac{\alpha \hbar}{\pi^2 \omega_0 l^2  \sqrt{8m}}.
\label{mopt}
\end {equation}
It follows from Eqs.(\ref{mac},\ref{mopt}) that for given $m$
$$ M \sim B^{3/2}, \:\:\: \Omega_c \sim B^{-1/2}.$$

The above results are in contrast with the usual
point of view \cite{las} that the motion of the electron
strongly  interacting with phonons in a
magnetic field is characterized by two frequencies only:
the frequency of the  fast
motion of the electron in the PW and the
 frequency of the  cyclotronic rotation of the polaron as a
whole with the mass not depending (or slightly depending)
on $B$. In fact this is true only when $\omega_c$
is small compared with the characteristic frequency of
the electron motion in the PW. Our results are valid in the
opposite case.

Note finally, that the splitting of the LL can be
apparently observed in Ge samples, where the deformation
potential is considerably large. Simple estimation shows
that in this material effective coupling constant $\alpha$ becomes
sufficiently large at $B > 50$ T.

\section*{Ascknowlegements}

The authors are
grateful to A.Alekseev, A.Barybin, and
M.Dyakonov for very useful
discussions. This work was supported in part by
the Swedish Royal Academy of Science (grant 1240)
 and by the
Russian Foundation for Basic Research (Grants 96-02-17894 a
and 96-02-17896).  Partial
support for one of the authors (V.Yu.K.) was provided
by a fellowship of INTAS Grant 93-2492-e within the research
program of International Center for Fundamental Physics in
Moscow.
This work was also supported by Grant 1001 within
the program "Physics of Solid State Nanostructures".

\widetext


\begin{thebibliography}{6}
\bibitem{stat.wave} Note, that here 
 we write the Hamiltonian in the
"stationary wave" representation which is different from
the one used in Ref.(\onlinecite{mi})
\end{thebibliography}
\end{document}